\documentclass[prb,aps,twocolumn,superscriptaddress,groupaddress]{revtex4-1} 
\usepackage{amsmath}  
\usepackage{amsfonts} 
\usepackage{graphicx} 
\usepackage{tikz}
\usepackage[siunitx]{circuitikz} 
\usetikzlibrary{shapes,arrows}
\usetikzlibrary{decorations.pathmorphing}
\usepackage{siunitx}
\usepackage{wasysym}
\usepackage{booktabs}
\usepackage[utf8]{inputenc}

\begin{document}

\title{GEM Foil Quality Assurance For The ALICE TPC Upgrade}

\author{Erik Br\"ucken}
\email{erik.brucken@iki.fi}
\affiliation{Helsinki Institute of Physics, P.O. Box 64, FIN-00014 University of Helsinki, Finland}
\author{Timo Hild\'en}
\affiliation{Helsinki Institute of Physics, P.O. Box 64, FIN-00014 University of Helsinki, Finland}

\begin{abstract}
The ALICE (A Large Ion Collider Experiment) experiment at the Large Hadron Collider (LHC) at CERN is dedicated to heavy ion physics to explore the structure of strongly interacting matter. 
The Time Projection Chamber (TPC) of ALICE is a tracking detector located in the central region of the experiment. It offers excellent tracking capabilities as well as particle identification. After the second long shutdown (LS2) the LHC will run at substantially higher luminosities. To be able to increase the data acquisition rate by a factor of 100, the ALICE TPC experiment has to replace the Multi-Wire Proportional Chamber (MWPC) –based readout chambers. The MWPC are operated with gating grid that limits the rate to $\mathcal{O}$\,(kHz). The new ReadOut Chamber (ROC) design is based on Gas Electron Multiplier (GEM) technology operating in continuous mode. The current GEM productions scheme foresees the production of more than 800 GEM foils of different types. To fulfill the requirements on the performance of the GEM TPC readout, necessitates thorough Quality Assurance (QA) measures. The QA scheme, developed by the ALICE collaboration, will be presented in detail.
\end{abstract}

\maketitle

\section{Introduction}
\label{intro}
ALICE at the LHC/CERN is dedicated to heavy ion physics, to explore the structure of strongly interacting matter. 

The LHC will be upgraded during the next long shutdown LS2 to run with higher collision rates. The ALICE experiment that currently runs with a recording event rate of 500~Hz will be upgraded to cope with an event rate of 50~kHz. To accomplish this, the ReadOut Chambers (ROC) of the present TPC, based on Multi-Wire Proportioanl Chambers (MWPC)~\cite{charpak} will be replaced by new ROCs utilizing Gas Electron Multiplier technologies (GEM)~\cite{sauli,ALICETPCUTDR}. 

The MWPC come with a huge ion back flow (IBF) that creates space charge distortions in the drift area. To avoid this a gating grid was introduced to eliminate the IBF with the disadvantage of having a limited readout rate of $\mathcal{O}$\,(kHz). GEM based readout chambers allow running in continuous mode while suppressing the IBF substantially. The remaining space charge distortions can be effectively corrected.
 
GEM detectors are a versatile and robust type of micropattern gaseous radiation detectors~\cite{sauli}. The active component is a thin (50~$\mu$m) polyimide foil coated with copper from both sides. Microscopic holes with typical sizes around 70~$\mu$m are etched chemically through the foil in a hexagonal pattern. 
When high voltage (HV) is applied over the copper electrodes, a high electrostatic field inside the holes, around 50~kV/cm, allows gas multiplication of electrons

\begin{figure}[!h]
\centering
\includegraphics[width=0.48\textwidth,clip]{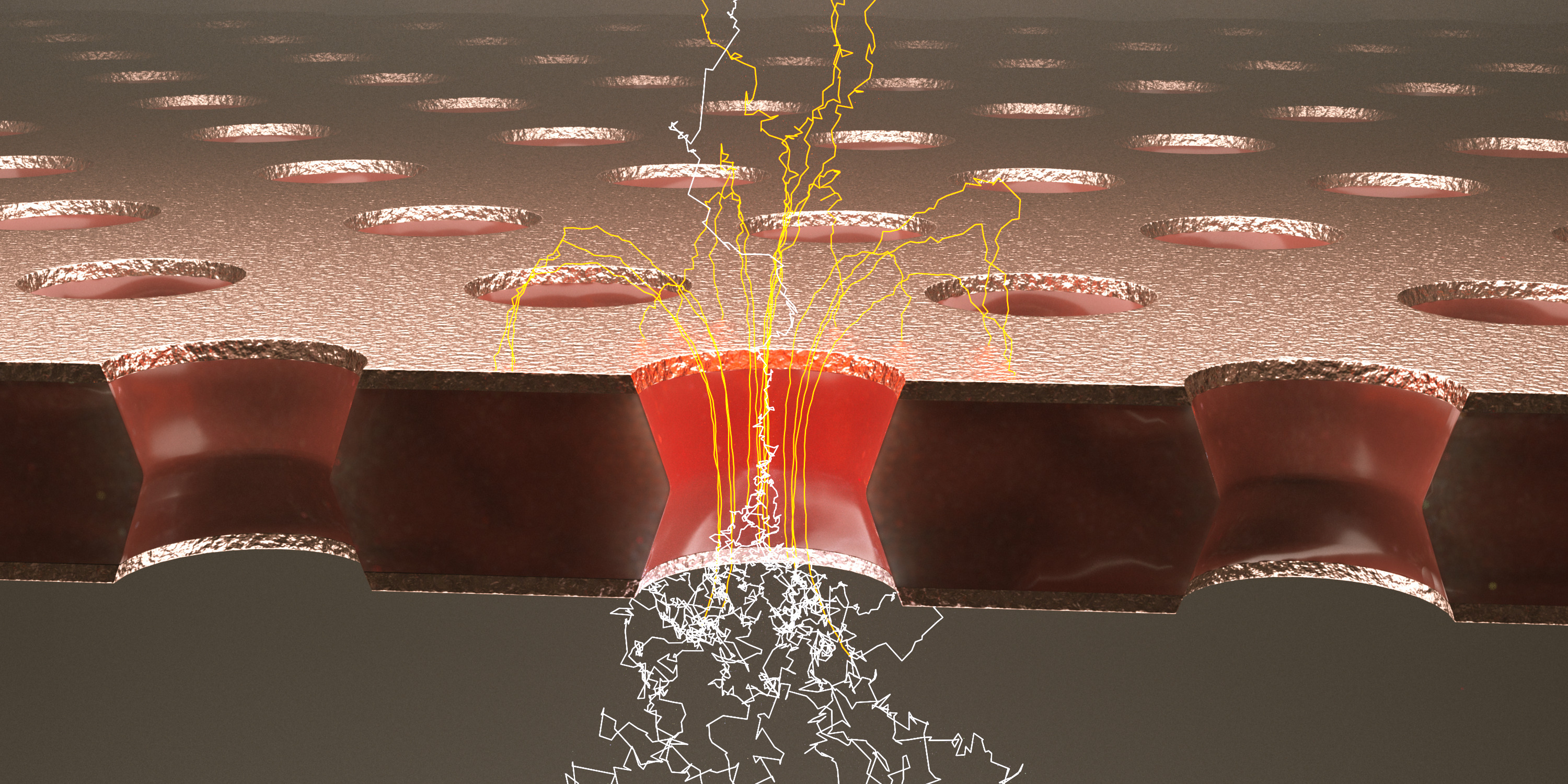}
\caption{Garfield simulation of an electron avalanche in a GEM foil.}
\label{fig-1}       
\end{figure}

Each hole in a GEM foil acts as an independent multiplication channel, enabling high resolution 2d detection of charged particles. Fig.\,\ref{fig-1} shows a magnified section of a GEM foil including a simulated electron avalanche by using \texttt{garfield++}~\cite{garfield}. A drifting electron entering from the top and the electron avalanche are shown in white. The back drifting positive ions are shown in yellow.

The ROC will be constructed with stack of 4 GEM foils that allow for continuous readout. The configuration can be tuned such to fulfill the design criteria, especially to maintain the present energy resolution and to have low ion feedback without the need for a gating grid. This could not be achieved with a stack of 3 GEM foils. More detailed information can be found in proceedings by Andreas Mathis~\cite{mathis}.

To fulfill the requirement of excellent detector performance, low IBF and stable operation, thorough quality assurance (QA) measures are of utmost importance. In this paper we focus solely on the QA of individual GEM foils but not on the QA of the ROCs.

\section{The ALICE TPC GEM foils}
\label{ALICEGEM}

The GEM foils for the ALICE TPC are processed with single mask technique by the Micro-Pattern Gaseous Detector (MPGD) workshop at CERN. The active area of a single TPC ROC, that is divided in inner and outer chambers (IROC, OROC), is approximately 0.89~m$^2$. The area is composed of 4 different GEM foils, one for the IROC and 3 for the OROC, simply because a single GEM foil of this size can not be produced so far. The TPC is composed of 18 ROCs on each side. One single ROC consists of 4 $\times$ 4 GEM foils. In total we get 576 single GEM foils that need to be tested or approximately 128~m$^2$. Due to an expected production yield of 85\% and the need for spare foils, an additional production of 25\% - 50\% is envisaged.

\section{Quality assurance scheme}
\label{QAM}

\noindent Motivation for the QA scheme follows from the experience of other experiments using GEM detectors for example COMPASS, TOTEM or KLOE with yields between 65-90\%. The rejection of unsuitable foils at earliest possible stage and selection of best foils in terms of gain uniformity, stability and number of defects is important. A quality monitoring after each production step is necessary.
The overall ROC production work flow, includes many production sites on two continents. Therefore defects have to be taken into account resulting from transport and foil handling at the different production places. 

Several methods have been developed within the ALICE TPC upgrade collaboration to test the GEM foils in all relevant aspects. 
All information and data collected by the different QA methods will be fed into a central database. Each foil will have a full history available for the ROC assembly and for a later forensic autopsy in case of detector failure.

The QA is divided into \textit{basic} QA and \textit{advanced} QA. The basic QA will be done directly at the production site of the GEM foils and consists of
\begin{itemize}
\item Cleaning procedure under HV,
\item Coarse optical inspection,
\item Leakage current measurement.
\end{itemize}

The advanced QA that will only be done at dedicated QA centers, consists of
\begin{itemize}
\item Advanced HV tests,
\item High resolution optical scanning,
\item Gain uniformity test.
\end{itemize}

A~\textit{traffic light system} is introduced to classify the usability of GEMs. In case the foil did not pass the basic production QA a \textit{red light} is given. 
A~\textit{yellow light} indicates that the foils passed the basic QA, i.e. no fatal defects, but did not show sufficient uniformity in gain, or shows non-optimal leakage currents. In case the foils passes all tests, a \textit{green light} is awarded. A continuous quality monitoring has to guarantee that the quality is maintained after each production step. 

\paragraph{HV cleaning procedure}
All HV sectors of the GEM foils are immediately ramped up to 600~V with a high current limit in the order of several tenth of $\mu$A. Through the sparks dust as well as chemical remnants from the production process are burned and therefore removed. Foils with several sparks occurring at the same position should be discarded as it could be a sign of defects in the GEM foils.

\paragraph{Coarse optical inspection}
Macroscopic defects of GEM foils should be spotted by eye and the help of a movable microscope. Such defects could be strong over-etching of holes, chemical remnants, deep scratches, etc. Foils holding such defects will be sent for re-cleaning or discarded.

\paragraph{Leakage current measurement and advanced HV tests} 
At production sites a leakage current test is foreseen. Each sector of the GEM foil will be ramped up in parallel to 500~V in humid air and leakage current will be measured using multi-channel picoammeters. For a good foil the leakage current should settle fast to under 500~pA. Only a small number of sparks, yet to be defined, will be allowed. At dedicated QA centers more advanced HV tests are planned, e.g. to measure the leakage current between neighboring HV sectors, to determine a quality factor and to perform long term HV tests. The HV test system will be introduced in more detail in Sec.\,\ref{HVTS}. 

\paragraph{High resolution optical scanning}
Each foil will be optically scanned with an automated microscopic camera system mounted onto an xyz-table. The system is able to measure the size and shape of each individual hole in a GEM foil as well as to detect microscopic defects such as over-etched holes, covered holes, etc. Detailed maps and histograms will be generated of each GEM foil showing geometrical properties and distribution of those. The individual selection criteria still need to be defined. The optical scanning system will be discussed in Sec.\,\ref{OSS}.

\paragraph{Gain uniformity test}
The gain uniformity is planned to be measured for at least one GEM foil of each type and for all GEM foils that exceed the limit of predicted gain uniformity derived from the high resolution scanning. The GEM foil will be mounted on top of a MWPC with a 2 dimensional readout. The local gain of the test foil will be measured by using the primary ionization above and below the foil using an $^{55}$Fe $X$-ray source. Foils should show a gain uniformity within 10\% RMS. The gain uniformity setup will be described in Sec.\,\ref{GUT}.

\section{HV test system}
\label{HVTS}

\begin{figure}[!t]
\centering
\includegraphics[width=0.48\textwidth,clip]{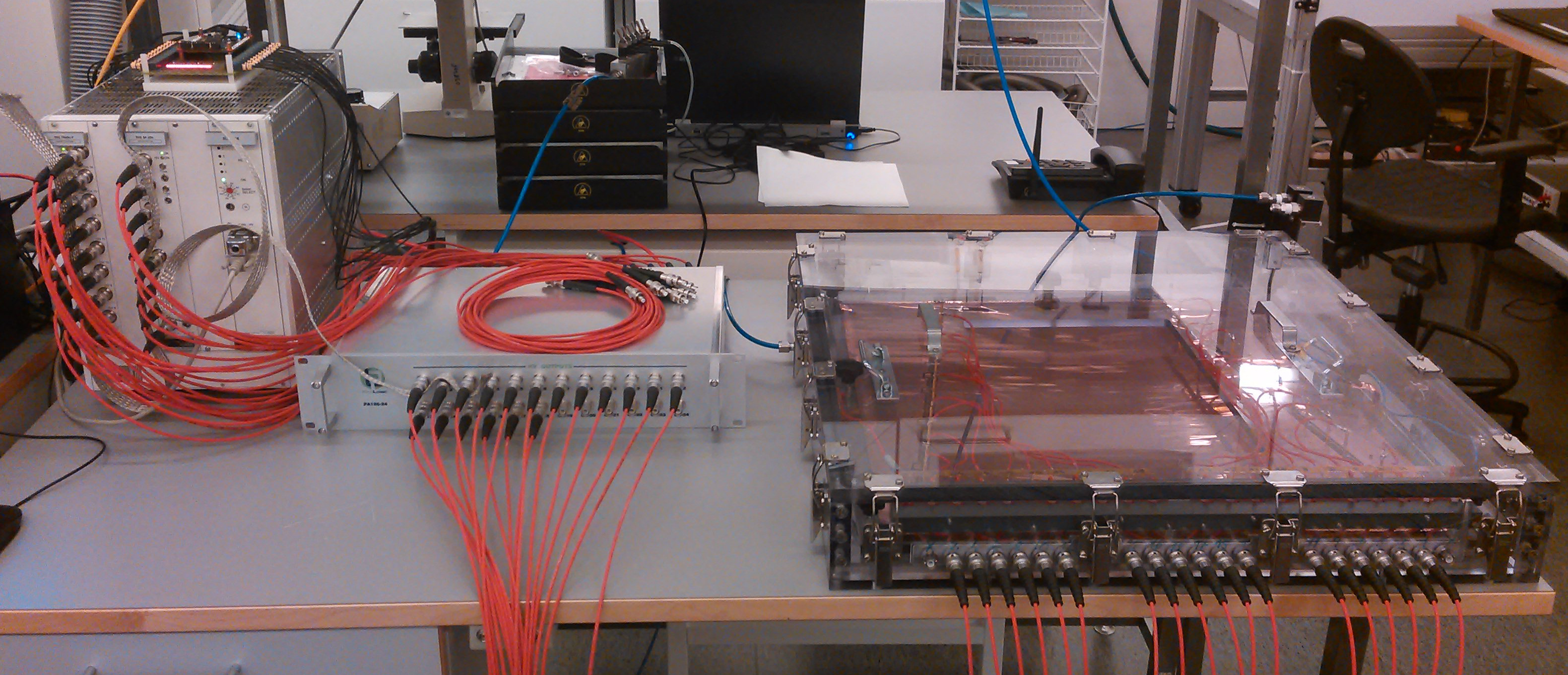}
\caption{HV test setup including HV power supply, picoammeters, HV box with gas flow and control pc.}
\label{fig-2}
\end{figure}

\begin{figure}[!htb]
\centering
\includegraphics[width=0.48\textwidth,clip]{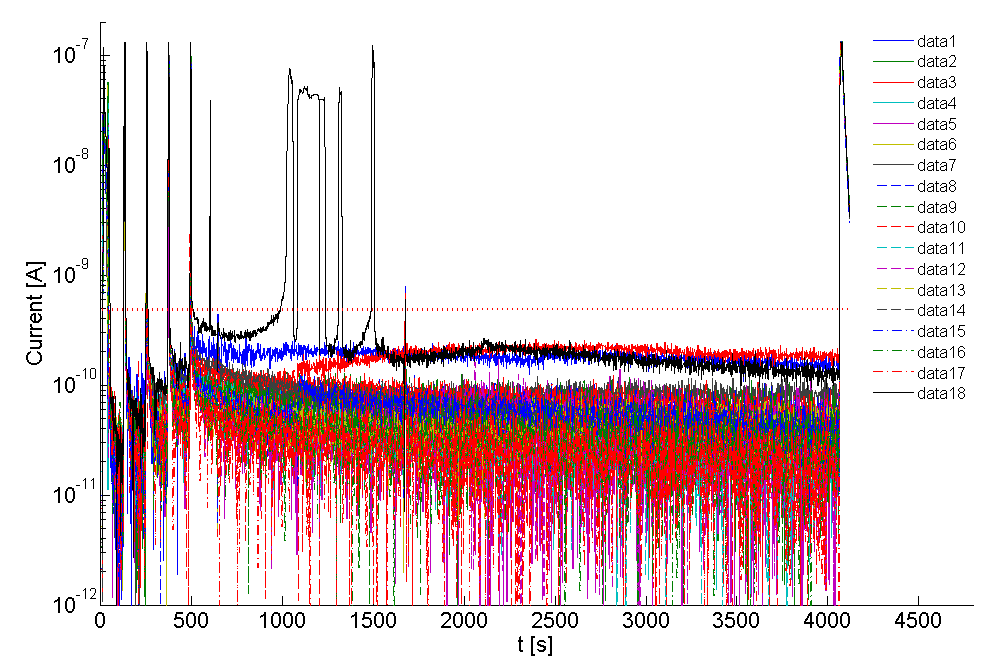}
\caption{Example leakage current measurement of an IROC test GEM foil.}
\label{fig-3}
\end{figure}

The HV test system built at HIP consists of a 24 channel high precision HV power supply~\cite{iseg}, a \textit{Picologic} 24 channel picoammeter~\cite{zagreb} and a Acrylic box with gas flow holding the GEM foil, as shown in Fig.\,\ref{fig-2}. The test system is controlled via custom made automated software based on Labview~\cite{labview}.
Scripts for certain test procedures can be implemented with data written out to a data file. As an example the results of an IROC GEM foil measurement using this HV test system is shown in Fig.\,\ref{fig-3}.  Each of the 18 HV sectors was ramped up in parallel in steps of 100~V every 60~s till 500~V was reached. The voltage was kept constant for 30~min before ramping it down. 
The criteria for the foil to pass this test is a stable leakage current below 500~pA for at least 30~min. A small number of random sparks right after reaching the voltage flattop are allowed. All sectors except except one passed the test. Channel 18 showed some non stable behavior with currents above 500~pA. The GEM foil was kept under constant nitrogen gas flow. 

The HV test system makes it possible to test interconnection currents between neighboring sectors of the foil using a checker board type of HV configuration. Also long term stability tests under HV are foreseen, where the foil is kept at 500~V in nitrogen atmosphere for about 12 hours. 

\section{High resolution optical scanning system}
\label{OSS}
\begin{figure}[]
\centering
\includegraphics[width=0.48\textwidth,clip]{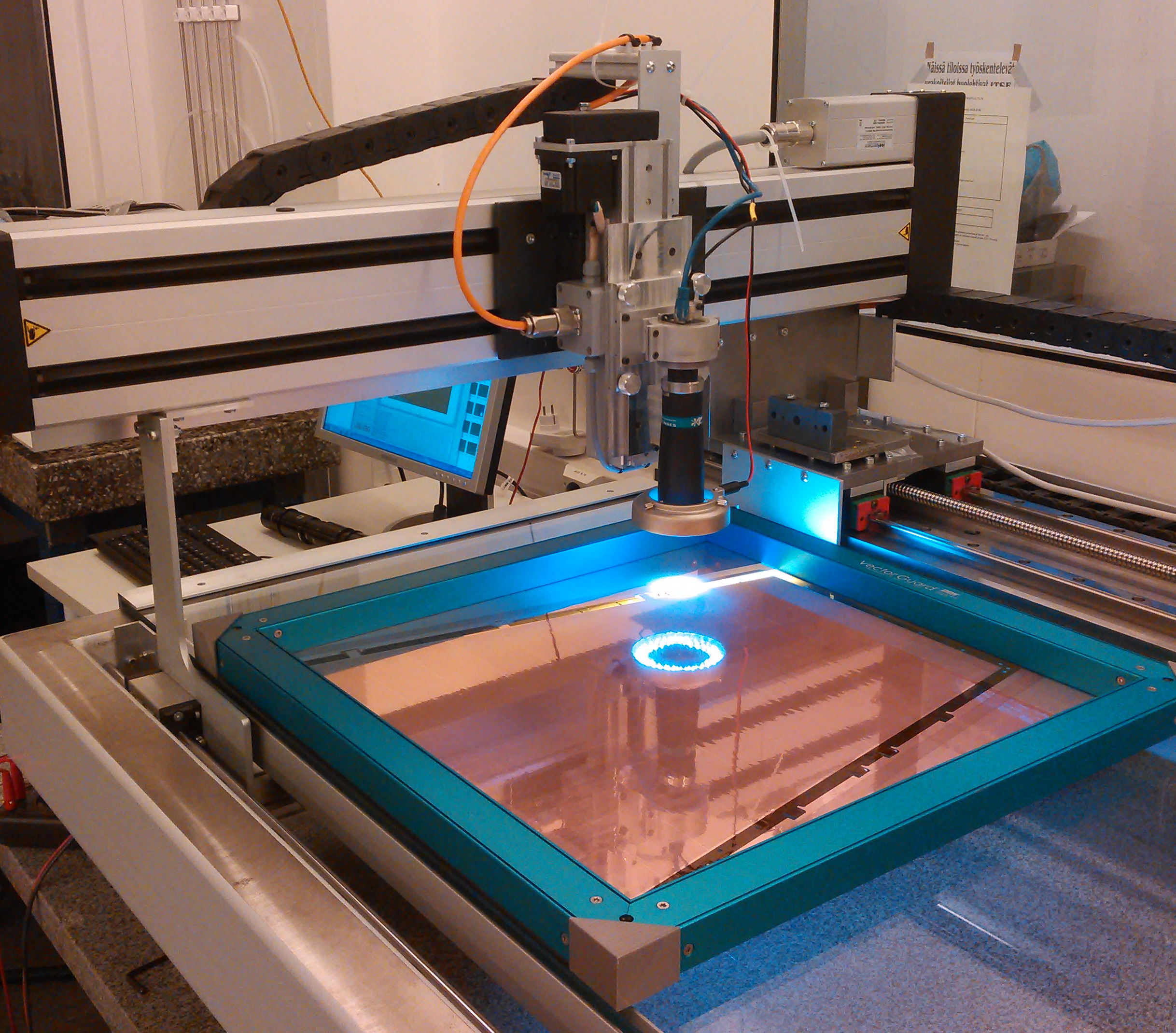}
\caption{High resolution optical scanning system working on an IROC GEM foil mounted inside a stretching frame.}
\label{fig-4}
\end{figure}

The optical scanner consists of a computer controlled xyz-robot on a glass table with background illumination, a high resolution camera and telecentric optics with coaxial lighting, shown in Fig.\,\ref{fig-4}. In addition a LED ring light is mounted. The installed 4.9 megapixel monochrome camera has a $1/2''$ cmos sensor with pixelsizes of 2.2 $\mu$m. The resolution depends on the quality and magnification of the available telecentric objectives. The system is controlled via custom made software based on Labview. 

\begin{figure}[]
\centering
\includegraphics[width=0.48\textwidth,clip]{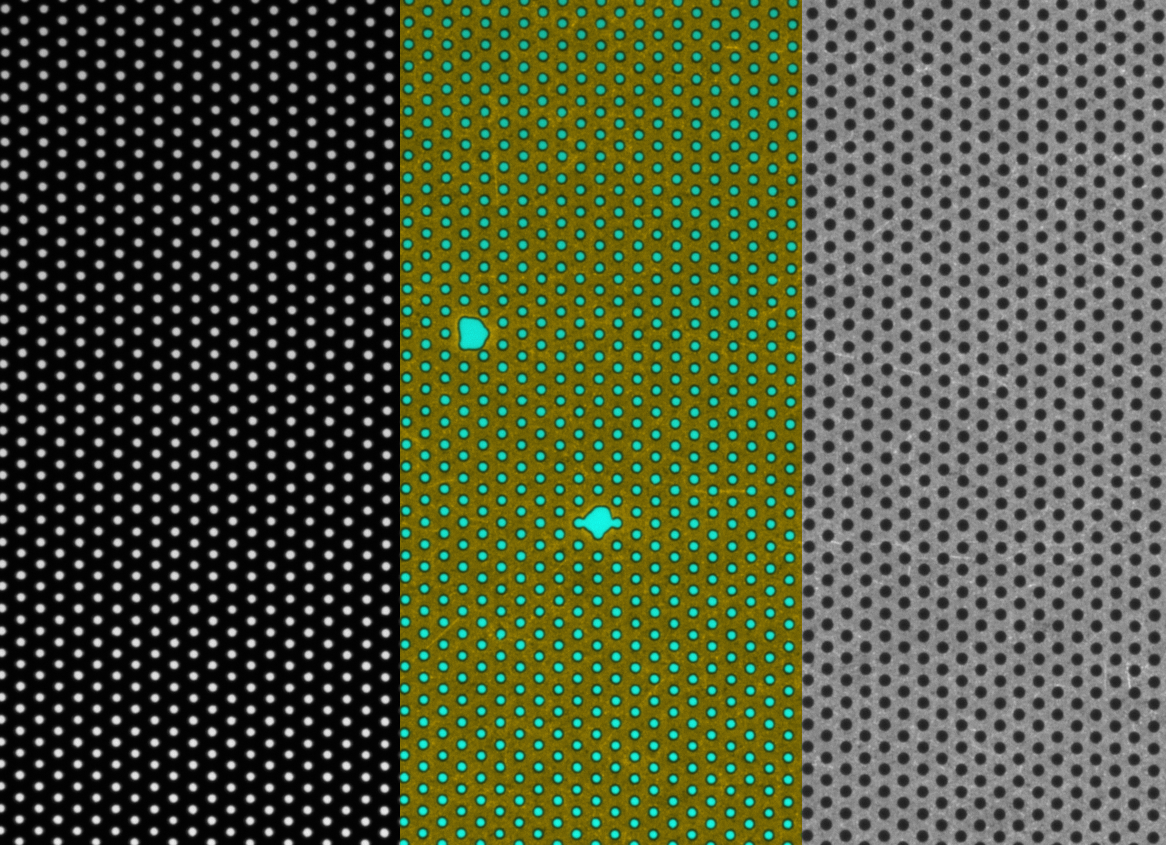}
\caption{Example image taken by the optical scanner in two exposure mode. Both exposures are shown including an overlay.}
\label{fig-5}
\end{figure}

The system takes images of the active area of GEM foils in either one or two exposure mode and transfers them to a network data-storage. The GEM foils are stretched before scanning to avoid image quality problems due to the telecentric optics. This also reduces the scanning time due to the reduction in refocusing of the camera. Presently the scanning time for an IROC takes about 14 hours with with a 1$\times$ magnification optics. An image example taken by the presented device is shown in Fig.\,\ref{fig-5}. The left side of the image shows the exposure with background light only, the right side the exposure only with foreground light and the center the overlay of both exposures in different color channels. In addition one can see two clear over-etching defects. 

The images are then analyzed offline using a custom made software described in detail in \cite{QAGEM_NIMA}. First the images are pre-processed mainly to enhance the contrast followed by segmentation using the edge detection Canny algorithm. The edge contours are then fitted by ellipses from which geometrical parameters are extracted. Classification methods based on neural networks are then used to determine whether the objects are inner or outer holes or defects.

\begin{figure}[]
\centering
\includegraphics[width=0.48\textwidth,clip]{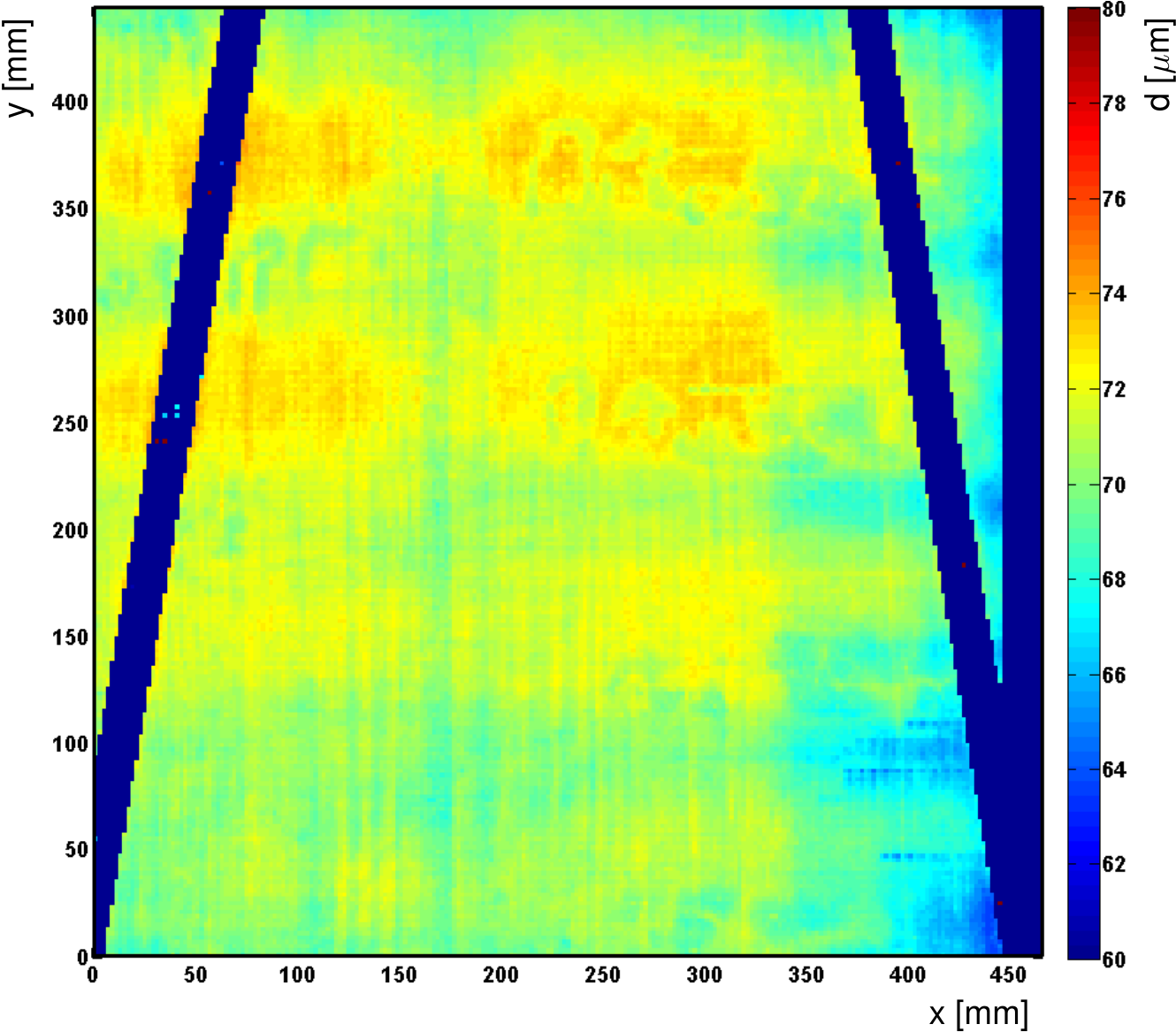}
\caption{Map of the outer hole diameters of an IROC GEM foil.}
\label{fig-6}
\end{figure}

The results are given in histograms and maps such as shown in Fig.\,\ref{fig-6}. Here, the map of the outer hole diameters of an IROC GEM foils is shown. The diameters vary between 64 to 74~$\mu$m (not calibrated). 
The criteria for a foil to pass the optical scanning are based on the uniformity of the hole properties and the number and type of defects. Detailed criteria are yet to be defined as hole size distributions from single mask etching technique have so far not been studied in detail due to the new production method.

\section{Gain uniformity test}
\label{GUT}

\begin{figure}[!h]
\centering
\includegraphics[width=0.48\textwidth,clip]{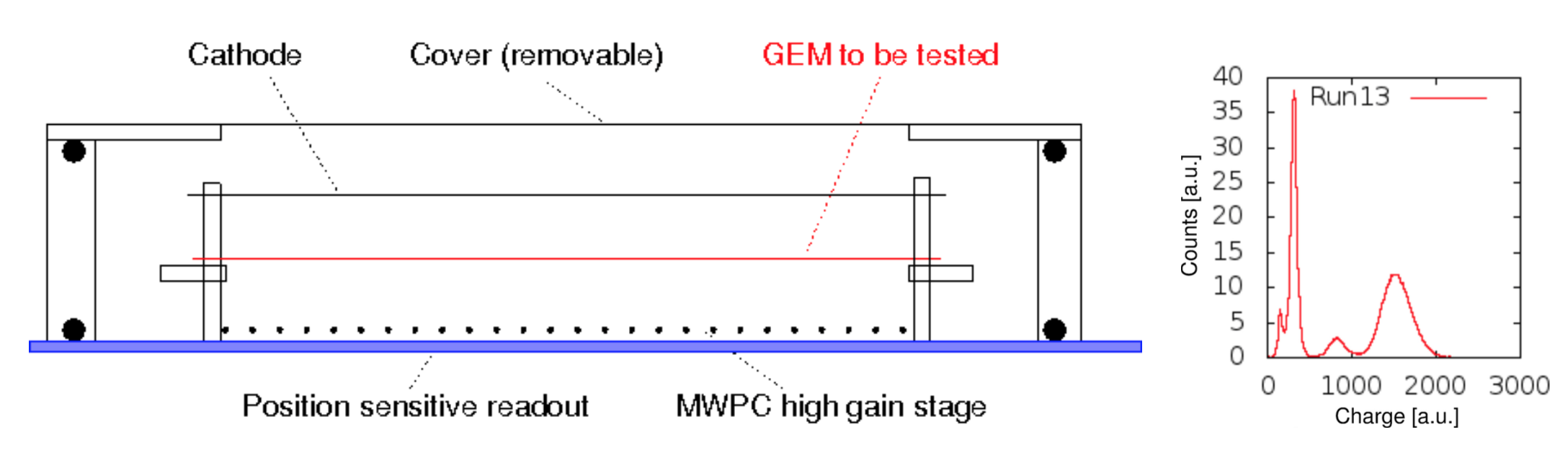}
\caption{Sketch and principle of the gain uniformity test system.}
\label{fig-7}
\end{figure}

The gain uniformity test system
 is based on a MWPC on top of a 2 dimensional readout with custom made electronics. A drawing that shows the construction idea of the system and the principle of the gain measurement is shown in Fig.\,\ref{fig-7}. 
The test GEM foil is installed on top of the MWPC and covered by a thin cathode.
Using an $^{55}$Fe $X$-ray source, that irradiates the full chamber, events with ionization points in the drift fields above and below the test GEM foil are recorded. The local relative gain is extracted as the ratio between both characteristic $^{55}$Fe peaks (Fig.\,\ref{fig-7} right). 

\begin{figure}[!h]
\centering
\includegraphics[width=0.48\textwidth,clip]{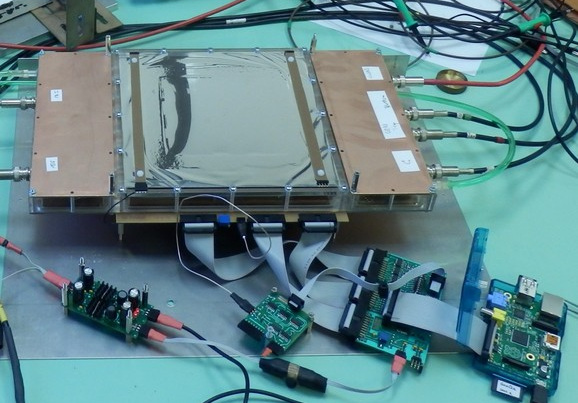}
\caption{Gain uniformity test system.}
\label{fig-8}
\end{figure}

It was shown that the local relative gain measurement is not affected by varying atmospheric conditions.
The prototype test system (Fig.\,\ref{fig-8}) is currently replaced by a full size system that can measure the largest ROC GEM foils. The custom electronics with a 12 bit ADC has enough dynamical bandwidth to measure also higher relative gains up to 30.

\section{Combination of tools}
\label{RD}

By comparing the results of the optical scanning, i.e. the individual geometrical properties of the holes, with the local gain measurements of the gain uniformity test system, a distinct correlation was found. 

A first successful result was shown in~\cite{QAGEM_NIMA}. This study has been successfully continued within the ALICE TPC upgrade collaboration. Preliminary results are shown in~\cite{timoposter}, where neural networks have been used to predict the gain of a GEM foil based entirely on geometrical foil and electric field properties. The training was done using measurements of several individual GEM foils. The GEM foil used for the prediction was not used in the training. We can conclude that the gain measurement and the prediction from optical scan data are in good agreement. However the method has so far only be used for foils produced with double mask technique and with limited statistic. Therefore it is planned to apply this method on a large data sample of GEM foils produced in single mask technique.

It is foreseen to use this knowledge to predict the gain uniformity to some extent using the geometrical properties only as it might not be possible to measure the relative gain of each individual foil.

\section{Present activity and conclusions}

Currently a GEM sample of 15 IROC foils is used to evaluate the QA methodology and to fine tune the QA protocol. The results will be used to fine tune the quality selection criteria such as a suitable leakage current per sector, hole size distribution, etc. 

The production phase is anticipated during the first quarter of 2016. The collaboration has built an extensive QA program for GEM foils. The infrastructure and method will be ready for the start of the production. Each individual GEM foil will have a full history record that is put into a database. The goal is to establish a general QA scheme that also solves many open questions on GEM QA R\&D. 

\section*{Acknowledgements}

We thank Dezso Varga from the Wigner RCP, Budapest, for providing us information about the gain mapping device, the ALICE team of TUM for providing the HV box and the non ALICE personnel of the Helsinki Institute of Physics laboratory that are involved in the infrastructure buildup.

\end{document}